\documentclass[aps,floats,twocolumn,showpacs]{revtex4-1}
\usepackage[cp1251]{inputenc}
\usepackage[english]{babel}
\usepackage{amssymb}
\usepackage{amsmath}
\usepackage{graphicx}
 \def\ds{\displaystyle}
 \def\bc{\begin{center}}          \def\ec{\end{center}}
 \allowdisplaybreaks

\begin{document}
 \title{Second harmonic electromagnetic emission of a turbulent magnetized plasma driven by a powerful electron beam}
 \author{I.V.\, Timofeev}
 \affiliation{Budker Institute of Nuclear Physics SB RAS, 630090, Novosibirsk, Russia \\
 Novosibirsk State University, 630090, Novosibirsk, Russia}
% \date{\today}
 \begin{abstract}
The power of second harmonic electromagnetic emission is calculated
for the case when strong plasma turbulence is excited by a powerful
electron beam in a magnetized plasma. It is shown that the simple
analytical model of strong plasma turbulence with the assumption of
a constant pump power is able to explain experimentally observed
bursts of electromagnetic radiation as a consequence of separate
collapse events. It is also found that the electromagnetic emission
power calculated for three-wave interaction processes occurring in
the long-wavelength part of turbulent spectrum is in
order-of-magnitude agreement with experimental results.
 \end{abstract}
 \maketitle

Electromagnetic emission of a turbulent plasma at the fundamental
plasma frequency $\omega_p$ and its harmonics has been the subject
of active theoretical and experimental research for several decades.
This radiation carries information about properties of plasma
turbulence, and registration of plasma emission is one of the most
efficient ways of studying  physical processes occurring in space
plasmas. That is why in most papers the problem of turbulent plasma
emission is considered  in the context of space phenomena such as
type III solar radio bursts \cite{ginz,gur,kru,grog,gold,rob,li} and
emissions of planet's magnetospheres \cite{gur2,cai}. Our interest
to this problem is motivated by laboratory experiments \cite{arz},
in which electromagnetic radiation at the doubled plasma frequency
is generated during injection of a high-current relativistic
electron beam into the plasma confined in the GOL-3 multimirror
trap. In contrast to previous beam-plasma experiments
\cite{ben,che,hop,ben2}, we study emission properties of rather hot
($T=$1-2 keV) plasmas in sufficiently strong magnetic fields
($\Omega=\omega_c/\omega_p\sim 1$, where $\omega_c$ is the electron
cyclotron frequency).

Several generation mechanisms of second harmonic plasma emission
exist. In the context of type III radio bursts, either weakly
turbulent coalescence of Langmuir waves $\ell +\ell\rightarrow t$
\cite{wil,li2}, or generation of electromagnetic waves by collapsing
caverns in strong plasma turbulence \cite{gold,fre,ak} are
discussed. Moreover, such electromagnetic waves can be produced due
to the Langmuir harmonic waves \cite{yo} scattering off density
fluctuations. It is obvious that the power of electromagnetic
emission depends essentially on what nonlinear processes form the
turbulent spectrum. In the theory of weak turbulence the important
role is played by the electrostatic Langmuir decay $\ell \rightarrow
\ell^{\prime}+s$ ($s$ denotes ion-acoustic waves) and the reverse
process $\ell +s \rightarrow \ell^{\prime}$. In an optically thick
plasma, formation of a turbulent spectrum can be also affected by
nonlinear processes $\ell \rightarrow t+s$ and $\ell +s \rightarrow
t$ involving electromagnetic waves with the plasma frequency
$\omega_p$. In models of strong plasma turbulence
\cite{gal,gold2,rob2} wave energy transfers through the spectrum due
to  Langmuir wave scattering off density fluctuations and collapse
of localized Langmuir wave packets. Alternative models of strong
plasma turbulence, in which wave collapse is supressed by either the
direct conversion of Langmuir waves to damping modes \cite{gal2} or
radiative losses \cite{mai}, are also discussed.

Experiments on turbulent plasma heating at the GOL-3 multimirror
trap show that intensity of electromagnetic emission demonstrates
not only smooth variation in time, but also bursts with the duration
of 2-10 ns, which we tend to associate with separate collapse
events. Thus, the model of strong plasma turbulence, proposed in
Ref. \cite{gal} and verified later by numerical simulations
\cite{deg}, seems most appropriate for our experiments. In Ref.
\cite{kru} this analytic model is used to estimate the power of
electromagnetic emission in the problem of type III radio bursts. In
order to explain the results of laboratory beam-plasma experiments
we will modify the model of Ref. \cite{deg} by taking into account
saturation of the pumping power due to beam trapping and generalize
the calculation procedure of the second harmonic emission power of
Ref. \cite{kru} to an arbitrary magnetic field.

According to the model of strong plasma turbulence \cite{gal,deg} an
isotropic turbulent spectrum of an unmagnetized plasma can be
divided into three typical regions: source region, inertial range
and dissipation region. The source region occupies small wavenumbers
$k<k_{M}\simeq \sqrt{W/(nT)}/r_{D}$ ($r_D$ is the Debye length) and
consists of untrapped Langmuir waves, which are the products of
beam-driven Langmuir waves scattering off long-wavelength density
fluctuations. It is assumed that the spectral density of wave energy
inside this part of the spectrum is independent on $k$. In the
inertial range, modulation instability results in trapping of
Langmuir waves in local density wells and is followed by the wave
collapse, which is responsible for the formation of the power-law
spectrum. In the small-wavelength region various dissipation
mechanisms come into force and spectral wave energy decreases
rapidly with the increase of $k$. Thus, second harmonic plasma
emission can be generated (i) due to Langmuir wave coalescence in
the source region containing most of the wave energy and (ii) due to
the collapse of trapped plasma oscillations which, despite the low
energy content, reach high energy densities at the late stage of
collapse and result in radiation bursts.

Let us calculate the level of the wave energy density $W$ in a
plasma turbulence that is pumped by an energy source with the power
$P$, and estimate the typical duration of radiation bursts produced
in separate collapse events. Energy balance between different parts
of turbulent spectrum can be written in the form
\begin{equation}\label{eq1}
    P\approx\omega_p \sqrt{\frac{\left\langle\delta
    n^2\right\rangle}{n^2}} W_R\approx \lambda(W)\omega_p
    \sqrt{\frac{m_e}{m_i}\frac{W}{nT}} W.
\end{equation}
The first equation describes the balance between pumping of
beam-excited resonant waves with the energy density $W_R$ and
dissipation produced by their scattering off long-wavelength density
fluctuations. The rms level of these fluctuations is determined by
the wave energy density $W$ concentrated in the source region:
$\sqrt{\left\langle\delta
    n^2\right\rangle}/n=\alpha W/nT$,
where $\alpha=0.7$ is the numerical coefficient obtained in 2D
simulations. The second equation in (\ref{eq1}) shows that the power
that comes to the source region from the pump is balanced by the
power that leaves this region due to the wave collapse. Here,
according to Ref. \cite{deg}, the rate of collapse is determined not
only by the rate of modulation instability, but also by the factor
$\lambda(W)=2\lambda (W/nT)^{1/2}$ accounting for finite time that
takes the collapse to reach the self-similar regime ($\lambda\simeq
0.7$).

The pump power is usually estimated as $P=2 \Gamma W_R$, where
$\Gamma$ is the linear growth rate of the beam--plasma instability.
In our experiments, however, the powerful electron beam relaxes in
the so-called trapping regime \cite{tim}, for which the energy pump
to the beam-excited coherent wave packets is saturated by the beam
nonlinearity. Indeed, 1D particle-in-cell simulations \cite{tim2}
show that in our case the pump power does not depend on evolving
parameters of the plasma turbulence. That is why in the balance
equation this power can be considered as a given constant
$P=\mbox{const}$. We can estimate this value from experimental data:
$P=\beta n T_0/\tau_0=100\, \mbox{kW/cm}^{3}$, assuming that it
takes $ \tau_0=3 \, \mu\mbox{s} $ to heat the plasma with the
density $n=2\cdot 10^{14} \, \mbox{cm}^{-3}$ up to the electron
temperature $T_0=1$ keV. The factor $\beta=6$ takes into account
that most of the dissipated wave energy goes to the formation of
high-energy tails.

For the resonant $W_R$ and nonresonant $W$ wave energies in this
model we get
\begin{align}\label{eq2}
    \frac{W}{nT}\approx &\frac{1}{\sqrt{2
    \lambda}}\left(\frac{m_i}{m_e}\right)^{1/4} \left(\frac{P}{\omega_p n
    T}\right)^{1/2}, \\
    &\frac{W_R}{W}\approx \frac{2
    \lambda}{\alpha}\sqrt{\frac{m_e}{m_i}}.
\end{align}
The characteristic duration of wave collapse can be estimated as
$$ \tau_c\sim \frac{1}{2 \lambda \omega_p} \sqrt{\frac{m_i}{m_e}} \frac{nT}{W}.$$
For typical parameters of beam-plasma experiments at the GOL-3
multimirror trap, the wave energy density reaches the value
$W/nT=0.01$ for the electron temperature  $T=1$ keV. The collapse
duration at the same stage appears to be 3-4 ns, which is in a good
agreement with the duration of radiation bursts observed
experimentally. Thus, experimental data do not contradict the basic
ideas of our model, and we can use estimates for the wave energy
density and the characteristic width of the energy-containing region
of plasma turbulence to calculate the emission power. We also assume
that these estimates are not extremely sensitive to the external
magnetic field.

Let us calculate the power of second harmonic plasma emission which
is generated due to coalescence of untrapped Langmuir waves $\ell
+\ell \rightarrow t$ in the long-wavelength part of turbulent
spectrum $k<k_M$. By Langmuir waves in the cold magnetized plasma we
mean plasma oscillations pertaining to the upper-hybrid branch.
\begin{figure*}[htb]
\bc\includegraphics[width=474bp]{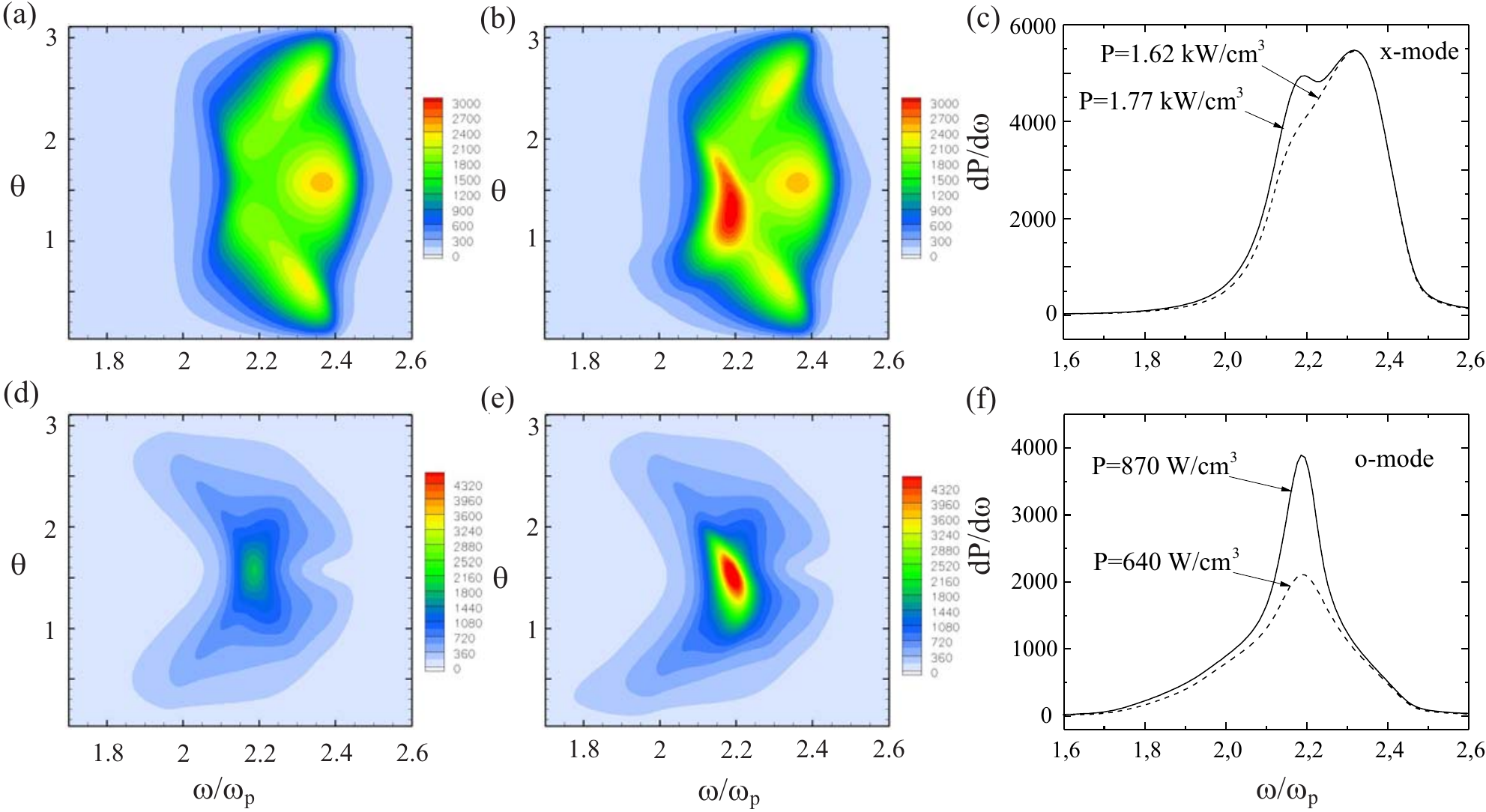} \ec \caption{ The
spectral power $dP/(d\omega d\theta)$ of x--mode emission for the
isotropic turbulent spectrum (a) and for the spectrum including
resonant waves (b). The spectral power $dP/(d\omega d\theta)$ of
o--mode emission for the isotropic turbulent spectrum (d) and for
the spectrum including resonant waves (e). The spectral power
$dP/d\omega$ of x--mode (c) and o--mode (f) emission for the
isotropic (dashed line) and anisotropic (solid line)
spectrum.}\label{r1}
\end{figure*}
Nonlinear interaction of such waves can be described in the
framework of weak turbulence, but in contrast to the standard theory
we will take into account model damping of two-time correlation
functions, which is used to describe the effect of finite life--time
of Langmuir plasmons due to their scattering off density
fluctuations. Let us represent the electric field in the following
form
$$ {\bf E}({\bf r},t)=\frac{1}{(2 \pi)^{3/2}}\sum\limits_{\sigma} \int {\bf e}^{\sigma}_k
E^{\sigma}_k(t) e^{\ds i {\bf k r}-i \omega^{\sigma}_k t } d^3 k,$$
where $\omega_k^{\sigma}$ and ${\bf e}_k^{\sigma}$ denote
eigenfrequencies and eigenvectors of linear plasma modes,
$E_k^{\sigma}$ --- their slowly varying amplitudes, and $\sigma$
indicates the branch to which they belong. In the general case,
three-wave interaction $\sigma^{\prime} + \sigma^{\prime\prime}
\rightarrow \sigma$ is described by the equation
\begin{equation}\label{eq4}
    \frac{\partial E_k^{\sigma}}{\partial t}=-\frac{4 \pi i e^{\ds i \omega_k^{\sigma} t}}{
    (\partial \Lambda^{\sigma} / \partial \omega)_{\omega_k^{\sigma}}}\frac{\partial}{\partial
    t} \left({\bf e}_k^{\ast \sigma}\cdot {\bf j}_k^{(2)}(t)\right),
\end{equation}
where $$\Lambda^{\sigma} ({\bf k},\omega)= |{\bf k}\cdot {\bf
e}^{\sigma}_k|^2 c^2-k^2 c^2+\omega^2 {\bf e}_k^{\ast \sigma}
{\bf\hat{\varepsilon}}({\bf k},\omega){\bf e}^{\sigma}_k,$$
$\hat{\varepsilon} ({\bf k},\omega)$ is the dielectric tensor and
${\bf j}_k^{(2)}$ is the Fourier transform of the second-order
nonlinear electron current, which in the cold plasma limit takes the
form
\begin{align}\label{eq5}
    \left({\bf j}_k^{(2)}\cdot{\bf e}_k^{\ast \sigma}\right) &=\frac{e n}{ (2
    \pi)^{3/2}} \int \frac{e^2 E^{\sigma^{\prime}}_{k_1} E^{\sigma^{\prime\prime}}_{k_2}}{m^2
    \omega^{\sigma^{\prime}}_{k_1}
    \omega^{\sigma^{\prime\prime}}_{k_2}} G^{\sigma \sigma^{\prime}
    \sigma^{\prime}}_{k, k_1, k_2} \times\nonumber \\ &\frac{e^{\ds -i \omega_{+}
t}}{\omega_{+}}  \delta({\bf k}-{\bf k}_1-{\bf k}_2) d^3 k_1 d^3
k_2, \\
G^{\sigma \sigma^{\prime} \sigma^{\prime}}_{k, k_1, k_2} =
&\omega_{+}
T_{k_2,\alpha\beta}^{\sigma^{\prime\prime}}e^{\sigma^{\prime\prime}}_{k_2,\beta}
T_{k_1, i j}^{\sigma^{\prime}} e^{\sigma^{\prime}}_{k_1, j}\times\nonumber \\
&\left(\frac{{k_1}_i}{\omega^{\sigma^{\prime}}_{k_1}} e_{k,
\alpha}^{\ast\sigma}+\frac{{k_2}_\alpha}{\omega^{\sigma^{\prime\prime}}_{k_2}}
e_{k, i}^{\ast \sigma}\right)+ e_{k, \alpha}^{\ast \sigma}
T_{\alpha\beta}^{(+)}g_{\beta}, \\
g_{\alpha}={k_2}_{\alpha}& T_{k_1, ij}^{\sigma^{\prime}} e_{k_2,
i}^{\sigma^{\prime\prime}} e_{k_1, j}^{\sigma^{\prime}}+ {k_2}_i
T_{k_1, ij}^{\sigma^{\prime}} e_{k_1, j}^{\sigma^{\prime}} \times\nonumber \\
&\left[T_{k_2, \alpha\beta}^{\sigma^{\prime\prime}} e_{k_2,
\beta}^{\sigma^{\prime\prime}}-\left(1-\frac{\Omega^2}{(\omega^{\sigma^{\prime\prime}}_{k_2})^2}\right)
e_{k_2, \alpha}^{\sigma^{\prime\prime}}\right]+ \nonumber
\\&(k_1,\sigma^{\prime} \leftrightarrow k_2,\sigma^{\prime\prime}),
\end{align}
\begin{align}
 T_{k, \alpha\beta}^{\sigma}=\frac{1}{1-\frac{\ds \Omega^2}{(\ds
\omega^{\sigma}_k)^2}}&\left[\delta_{\alpha\beta}-
i\frac{\Omega}{\omega^{\sigma}_k}e_{\alpha\beta\gamma}h_{\gamma}-\frac{\Omega^2}{(\omega^{\sigma}_k)^2}
h_{\alpha}h_{\beta}\right],\nonumber \\
&\omega_{+}
=\omega^{\sigma^{\prime}}_{k_1}+\omega^{\sigma^{\prime\prime}}_{k_2},\nonumber
\end{align}

In dimensionless units $\omega_p t$, $\omega/\omega_p$, $x
\omega_p/c$, $kc/\omega_p$, $eE_k^{\sigma} (\omega_p/c)^3/(m c
\omega_p)$ for time, frequency, position, wavenumber and electric
field amplitude of a plasma mode, respectively, three-wave
interaction processes $\ell +\ell \rightarrow t$ between Langmuir
and electromagnetic waves are described by the equation
\begin{multline}\label{eq6}
    \frac{\partial E_k^t}{\partial t}=-\frac{1}{2 (2 \pi)^{3/2}
    (\partial \Lambda^{\ell}/\partial \omega)_{\omega_k^t}}\int
    \frac{E_{k_1}^{\ell} E_{k_2}^{\ell}}{\omega_{k_1}^{\ell} \omega_{k_2}^{\ell}} G^{t \ell\ell}_{k, k_1, k_2}\times \\ e^{ i
    (\omega_k^t-\omega_{k_1}^{\ell}-\omega_{k_2}^{\ell})t} \delta({\bf k}-{\bf k}_1-{\bf k}_2) d^3 k_1 d^3
    k_2.
\end{multline}

According to Ref. \cite{kru} we assume that Langmuir waves,
scattering off long-wavelength density fluctuations, change their
phases stochastically with the characteristic frequency
$\nu/\omega_p= W^{\ell}/nT$. In this case, the temporal correlation
function of Langmuir electric fields can be written in the form
\begin{equation}\label{eq7}
    \left\langle E_{k}^{\ell}(t) E_{q}^{\ast\ell
    }(t^{\prime})\right\rangle = I_k^{\ell} \  \delta({\bf k}-{\bf
    q}) e^{\ds -\nu |t-t^{\prime}|}.
\end{equation}
For the average energy of Langmuir turbulence we get
\begin{align}\label{eq8}
    &\frac{W^{\ell}}{n m c^2}=\int W_k^{\ell} d^3 k, \\
    W_k^{\ell}=&\frac{1}{2 (2 \pi)^3 \omega_k^{\ell}} \left(\frac{\partial
\Lambda^{\ell}}{\partial \omega}\right)_{\omega_k^{\ell}}
I_k^{\ell}. \nonumber
\end{align}
The spectral energy density of electromagnetic waves produced in
spontaneous processes $\ell +\ell \rightarrow t$ is governed by the
equation
\begin{multline}\label{eq10}
\frac{\partial W_k^t}{\partial t}= \frac{2 \pi}{\omega_k^t (\partial
\Lambda^t /\partial \omega)_{\omega_k^t}}\times \\ \int
\frac{W_{k_1}^{\ell}
 W_{k_2}^{\ell} \left|G^{t\ell\ell}_{k, k_1, k_2}\right|^2 \Delta_{k, k_1, k_2} \delta({\bf k}-{\bf k}_1-{\bf k}_2)}{\omega_{k_1}^{\ell} (\partial
\Lambda^{\ell} /\partial \omega)_{\omega_{k_1}^{\ell}}
\omega_{k_2}^{\ell} (\partial \Lambda^{\ell} /\partial
\omega)_{\omega_{k_2}^{\ell}}} d^3 k_1 d^3 k_2,
\end{multline}
where $\Delta_{k, k_1, k_2}$ is the function describing correlation
broadening of the resonance $\omega_k^t-\omega_{k_1}^{\ell}
-\omega_{k_2}^{\ell}=0$:
$$ \Delta_{k, k_1, k_2}=\frac{2 \nu /\pi}{\left(\omega_k^t-\omega_{k_1}^{\ell} -\omega_{k_2}^{\ell}\right)^2+4 \nu^2}.$$
Thus, in the case of azimuthally symmetric turbulence, the spectral
power of second harmonic electromagnetic emission in units of
$nmc^2$ is given by the integral
\begin{equation}\label{eq11}
    \frac{d P}{d \omega} =  2 \pi \int\limits_0^{\pi} \sin \theta d\theta \left(\frac{k^2}{d\omega/dk}\frac{\partial W_k^t}{\partial
    t}\right)_{k(\omega)},
\end{equation}
where $k(\omega)$ is the solution of $\omega=\omega_k^t$ and
$\theta$ is the polar angle of ${\bf k}$.

Let us compute the spectral power of electromagnetic emission for
parameters typical for beam-plasma experiments in the GOL-3
multimirror trap. In the regime with the plasma density $n=2\cdot
10^{14} \mbox{cm}^{-3}$, the external magnetic field $\Omega=0.8$
and the electron temperature $T=1-2$ keV, the power of
electromagnetic emission was estimated experimentally as $0.1\div
1\, \mbox{kW/cm}^3$. In our theoretical model, the isotropic part of
long-wavelength plasma turbulence should occupy the spectral region
$k\in (0.1; 2.45) \omega_p/c$ and should contain the energy
$W^{\ell}/nT=0.01$ for the typical electron temperature $T=1$ keV.
Moreover, in the beam-excited turbulence there is an anisotropic
population of resonant Langmuir waves containing about 5\% of
turbulence energy. This part of energy is concentrated in a rather
small spectral region: $k\in (1.1; 1.3)$ and $\theta\in (0; 0.3)$.
The computation results for the emission power of ordinary (o--mode)
and extraordinary (x--mode) electromagnetic waves are presented in
Fig. \ref{r1}. To analyze the contribution of resonant waves to
plasma emission, we also present computations accounting for the
isotropic part of turbulence only.

The angular distribution of the emission power $dP/d\omega d\theta$
shows that both x--mode [Fig. \ref{r1} (a), (b)] and o--mode [Fig.
\ref{r1} (d), (e)] are radiated predominantly in the transverse to
the magnetic field direction. It is also shown in Fig. \ref{r1}(c)
and \ref{r1}(f) that the total emission power integrated over angle
and frequency reaches the value of 2 $\mbox{kW/cm}^3$ and is
dominated by the x--mode contribution. One can see that resonant
waves do not substantially affect x--mode emission and result in the
significant increase of the o--mode emission power. From the
emission spectrum it is also seen that the main role in generation
of electromagnetic waves is played by almost potential Langmuir
waves with frequencies $\omega>\omega_p$.

In conclusion, we calculate second harmonic electromagnetic emission
of a turbulent magnetized plasma driven by a powerful electron beam.
We found that the simple analytical model of strong plasma
turbulence with the assumption of constant pump power explains the
results of laboratory beam--plasma experiments at the GOL-3
multimirror trap. We show that the typical duration of
electromagnetic bursts observed in these experiments is in a good
agreement with the theoretical estimate of collapse duration. Our
theory does also predict by order of magnitude experimental results
for the total emission power and explains polarization of this
emission.

The author thanks Prof. A.V. Arzhannikov and Prof. I.A. Kotelnikov
for useful discussions. This work is supported by President grant
NSh-5118.2012.2,  grant 11.G34.31.0033 of the Russian Federation
Government and RFBR grants 11-02-00563-a, 11-01-00249-a.


\begin{thebibliography}{10}

\bibitem{ginz}
V.L. Ginzburg, V.V. Zheleznyakov, Sov. Astron. {\bf 2}, 235 (1958)

\bibitem{gur}
D.A. Gurnett, R.R. Anderson, Science {\bf 194}, 1159 (1976).

\bibitem{kru}
E.N. Kruchina, R.Z. Sagdeev, V.D. Shapiro, JETP Lett. {\bf 32}, 419
(1980).

\bibitem{grog}
R.J.M. Grognard, Sol. Phys. {\bf 81}, 173 (1982).

\bibitem{gold}
M.V. Goldman, Sol. Phys. {\bf 89}, 403 (1983).

\bibitem{rob}
P.A. Robinson, I.H. Cairns, and A.J. Willes, Astrophys. J. {\bf
422}, 870 (1994).

\bibitem{li}
B. Li, P.A. Robinson, I.H. Cairns, Phys. Plasmas {\bf 13}, 092902
(2006).


\bibitem{gur2}
D.A. Gurnett, S.D. Shawhan, and R.R. Shaw, J. Geophys. Res. {\bf
88}, 329 (1983).

\bibitem{cai}
I.H. Cairns and J.D. Menietti, J. Geophys. Res. {\bf 106}, 29515
(2001).

\bibitem{arz}
A.V. Arzhannikov, A.V. Burdakov, S.A. Kuznetsov, M.A. Makarov, K.I.
Mekler, V.V. Postupaev, A.F. Rovenskikh, S.L. Sinitsky, V.F.
Sklyarov,  Fusion Sci. and Technol. {\bf 59} (1T), 74 (2011).

\bibitem{ben}
G. Benford, D. Tzach, K. Kato, D.F. Smith, Phys. Rev. Lett. {\bf
45}, 1182 (1980).

\bibitem{che}
P.Y. Cheung, A.Y. Wong, C.B. Darrow, and S.J. Qian, Phys. Rev. Lett.
{\bf 48}, 1348 (1982).

\bibitem{hop}
H.J. Hopman and G.C.A.M. Janssen, Phys. Rev. Lett. {\bf 52}, 1613
(1984).

\bibitem{ben2}
A. B. Baranga, G. Benford, D. Tzach, K. Kato, Phys. Rev. Lett. {\bf
54}, 1377 (1985).



\bibitem{wil}
A.J. Willes, P.A. Robinson, and D.B. Melrose, Phys. Plasmas {\bf 3},
149 (1996).

\bibitem{li2}
B. Li, A.J. Willes, P.A. Robinson, and I.H. Cairns, Phys. Plasmas
{\bf 12}, 012103 (2005).


\bibitem{fre}
H.P. Freund and  K. Papadopoulos, Phys. Fluids {\bf 23}, 732 (1980);
 Phys. Fluids {\bf 23}, 1546 (1980).

\bibitem{ak}
K. Akimoto, H.L. Rowland, and K. Papadopoulos, Phys. Fluids {\bf
31}, 2185 (1988).

\bibitem{yo}
P.H. Yoon, R. Gaelzer, T. Umeda, Y. Omura, and H. Matsumoto, Phys.
Plasmas {\bf 10}, 364 (2003).

\bibitem{cyt}
V.N. Tsytovich, Theory of Turbulent Plasma (Consultants Bureau, New
York, 1977).


\bibitem{gal}
A.A. Galeev, R.Z. Sagdeev, V.D. Shapiro, and V.I. Shevchenko, Sov.
Phys. JETP {\bf 46}, 711 (1977).

\bibitem{gold2}
M.V. Goldman, Rev. Mod. Phys. {\bf 56}, 709 (1984).

\bibitem{rob2}
P. A. Robinson, Rev. Mod. Phys. {\bf 69}, 507 (1997).

\bibitem{gal2}
A.A. Galeev, R.Z. Sagdeev, V.D. Shapiro, V.I. Shevchenko, JETP Lett.
{\bf 24}, 21 (1976).

\bibitem{mai}
W. Main and G. Benford, Phys. Fluids B {\bf 1}, 2479 (1989).

\bibitem{deg}
L.M. Degtyarev, R.Z. Sagdeev, G.I. Solov'ev, V.D. Shapiro, and V.I.
Shevchenko, Sov. Phys. JETP {\bf 68}, 975 (1989).

\bibitem{tim}
I.V. Timofeev, K.V. Lotov,  Phys. Plasmas {\bf 13}, 062312 (2006).

\bibitem{tim2}
I.V. Timofeev, A.V. Terekhov,  Phys. Plasmas {\bf 17}, 083111
(2010).

\end{thebibliography}
\end{document}